\documentclass[prl,aps,twocolumn,showpacs,bibnotes,epsf]{revtex4}
\usepackage{graphicx, verbatim}
\newcommand{\Journal}[4]{#1 \textbf{#2}, #3 (#4)}
\usepackage{epstopdf}
\usepackage{CJK}
\usepackage[utf8]{inputenc}
\usepackage[english]{babel}
\usepackage{hyperref}
\hypersetup{colorlinks=true,linkcolor=blue,filecolor=magenta,urlcolor=cyan}

\begin{document}
\begin{CJK*}{GB}{gbsn}
\title{Dynamical Skyrmion State in a Spin Current Nano-Oscillator with Perpendicular Magnetic Anisotropy}
\author{R. H. Liu(ÁõÈÙ»ª)} \email[]{rhliu.phy@gmail.com}
\author{W. L. Lim}
\author{S. Urazhdin}\email[]{sergei.urazhdin@emory.edu}
\affiliation{Department of Physics, Emory University, Atlanta, Georgia 30322, USA}

\begin{abstract}
We study the spectral characteristics of spin current nano-oscillators based on the Pt/[Co/Ni] magnetic multilayer with perpendicular magnetic anisotropy. By varying the applied magnetic field and current, both localized and propagating spin wave modes of the oscillation are achieved. At small fields, we observe an abrupt onset of the modulation sidebands. We use micromagnetic simulations to identify this state as a dynamical magnetic skyrmion stabilized in the active device region by spin current injection, whose current-induced dynamics is accompanied by the gyrotropic motion of the core due to the skew deflection. Our results demonstrate a practical route for controllable skyrmion manipulation by spin current in magnetic thin films.
\end{abstract}


\maketitle
\end{CJK*}

The possibility to induce dynamical states of nanomagnets or change their static configuration by the current-induced spin torque (ST)~\cite{slon1,berger} has stimulated intense research into current-induced phenomena in magnetic systems. While the early experiments utilized spin-polarized electric currents in magnetic multilayers~\cite{tsoiprl,cornellorig}, recent studies focused on the effects of spin current produced due to the spin-orbit interaction (SOI) in bilayers of ferromagnets (F) with heavy nonmagnetic metals (N)~\cite{miron,liuprl,Ando_SHE,wang}. The spin-orbit effects that contribute to the current-induced phenomena include the spin Hall effect (SHE) originating from SOI in N, the Rashba effect~\cite{rashba} and the Dzyaloshinskii-Moriya interaction (DMI)~\cite{dmi}, both originating from the broken inversion symmetry at the F/N interface. The DMI produces chiral effective fields that can result in nontrivial topological magnetic structures such as spirals, skyrmions and chiral domain walls~\cite{Braun}. The possibility to control these textures by spin current in F/N structures has been extensively theoretically analyzed~\cite{obler, sampaio, zhou}, but few studies have investigated this possibility experimentally~\cite{bode,romming}.

Here, we report an experimental study of the dynamics in a F/N bilayer with perpendicular magnetic anisotropy (PMA) induced by a locally injected spin current. Both localized and propagating-mode dynamical regimes of magnetization can be achieved by varying the applied magnetic field and the current. At small fields, we observe modulated spectral peaks that are reminiscent of the ``droplet" soliton dynamical mode~\cite{hoefer1} recently observed in the conventional magnetic multilayer spin torque nano-oscillator (STNO) at large out-of-plane fields~\cite{mohseni}. Micromagnetic simulations confirm that the droplet-like spectra are associated with the dynamics of a topologically nontrivial nanoscale magnetic bubble (or equivalently a dynamical skyrmion)~\cite{bubble} stabilized by the local spin current injection, suggesting a route for the controllable manipulation of nontrivial spin textures in magnetic films.


\begin{figure}[htbp]
\centering
\includegraphics[width=0.35\textwidth]{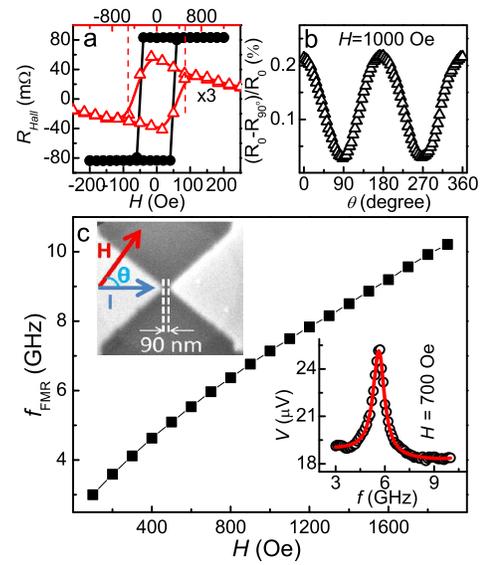}
\caption{(Color online) Magnetic characteristics of the Pt(5)/[Co(0.2)/Ni(0.3)]$_6$ film at $295$~K. (a) Anomalous Hall effect measured in a film patterned into a 4 $\times$ 14~$\mu$m Hall cross, with in-plane (triangles) and out-of-plane (circles) field. (b) Dependence of SCAO resistance on the direction of in-plane field $H=1$~kOe, due to AMR of the magnetic film. (c) $f_{FMR}$ vs in-plane field measured in the SCAO. Top inset: scanning electron micrograph of the device and the experimental layout. Bottom inset: ST-FMR voltage vs frequency $f_{ex}$ of the applied microwave current $I_{ac}=2$~mA, at $H=700$~Oe~\cite{Iac}. The curve is the best fit with a sum of symmetric and antisymmetric Lorentzians, accounting for the effects of ST and the Oersted field.}\label{fig1}
\end{figure}

The studied device is based on a Pt(5)/[Co(0.2)/Ni(0.3)]$_6$ magnetic multilayer with PMA~\cite{Daalderop}. Thicknesses are given in nanometers. The film was deposited on the sapphire substrate by magnetron sputtering at room temperature. The magnetic properties of the multilayer were characterized by measurements of the anomalous Hall effect [Fig.~\ref{fig1}(a)]. A square hysteresis loop obtained with field perpendicular to the film plane indicates that the magnetic film has a well-defined PMA. The hysteresis loop measured with the field oriented at $5^\circ$ relative to the film plane is more rounded and broader, and the maximum Hall resistance achieved at small $H$ is significantly smaller. These behaviors are consistent with the magnetization reversal by gradual nucleation and growth of magnetic bubbles, as indicated by our micromagnetic simulations~\cite{simububble} and directly demonstrated for similar systems by magnetic imaging~\cite{gubbiotti, chen}. The hysteresis loop closes above $H_0=500$~Oe, which can be identified with the disappearance of the bubble domains. We will show below that a transition between different magnetization oscillation modes of the nanodevice occurs precisely at $H_0$.

Further characterization was performed by magnetoelectronic measurements of the actual device, which consisted of the Pt/[Co/Ni] multilayer patterned into a disk with diameter of 4~$\mu$m and two pointed Au(100) electrodes separated by a $90$~nm gap that were fabricated on top of the disk [top inset in Fig.~\ref{fig1}(c)]. By applying a current between the electrodes, a spin current is locally injected into the [Co/Ni] layer due to a combination of the spin Hall effect in Pt~\cite{Ando_SHE, wang, liuprl} and Rashba effect at the Pt/[Co/Ni] interface~\cite{miron, rashba}, resulting in magnetization auto-oscillation. We label our device the spin current auto-oscillator (SCAO). The magnetization state of SCAO was detected using the dependence of the device resistance $R$ on the angle between the magnetization and the direction of the current flow, which exhibited a $180^\circ$ periodicity consistent with the anisotropic magnetoresistance (AMR) of the [Co/Ni] multilayer~\cite{mcguire} [Fig.~\ref{fig1}(b)]. To determine the dynamical characteristics of the [Co/Ni] layer, the ferromagnetic resonance (FMR) frequency $f_{FMR}$ was measured by the ST-FMR technique~\cite{mosendz} [Fig.~\ref{fig1}(c)]. All the measurements described below were performed with $H$ tilted by $5^\circ$ relative to the film plane and at $\theta=60^\circ$, which enabled generation of microwave voltages at the first harmonic of the magnetization oscillation frequency. Spectroscopic measurements were performed at $T=140$~K, and the results were confirmed for three devices.

\begin{figure}[htbp]
\centering
\includegraphics[width=0.40\textwidth]{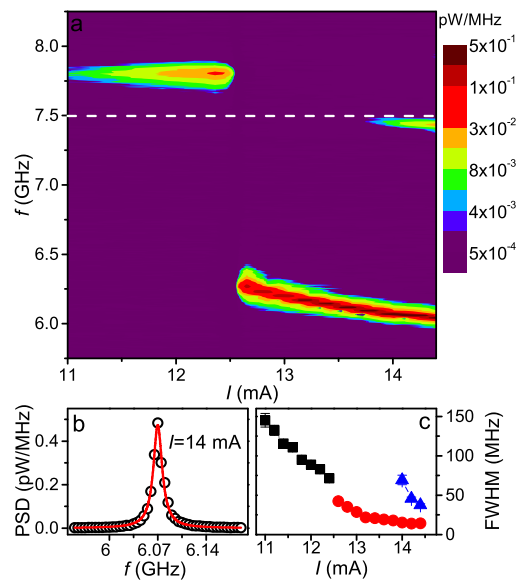}
\caption{(Color online). Dependence of the microwave generation characteristics on current $I$ at $H=1.1$~kOe. (a) Psudo-color plot of the spectra obtained at $I$ between $11$~mA and $14.4$~mA increased in $0.2$~mA steps. (b) Symbols: power spectral density (PSD) at $I$=14 mA. The linewidth was determined by fitting the spectra with the Lorentzian function. (c) Dependence of the FWHM on current for the propagating mode at small $I<12.5$~mA (squares), bullet mode at $I>12.5$~mA (circles), and the quasi-propagating mode at $I>14$~mA (triangles).}\label{fig2}
\end{figure}

Magnetization oscillation was observed when a sufficiently large dc current $I$ was applied to the device, at fields ranging from below $100$~Oe to almost $2$~kOe. Figure~\ref{fig2} shows the dependence of the spectral characteristics on $I$ at $H=1.1$~kOe. The oscillation frequency $f=7.8$~GHz at the onset current $I_C = 11$~mA, was above f$_{FMR}=7.5$~GHz obtained from ST-FMR measurements [Fig.~\ref{fig1}(c)], and exhibited a very weak blueshift with increasing current. These characteristics are consistent with Slonczewski's quasi-linear theory of propagating spin wave emission by STNO~\cite{slon1}.

The oscillation characteristics abruptly changed above the threshold current $I_t=12.6$~mA. The frequency  dropped to $f=6.3$~GHz, which is far below $f_{FMR}=7.5$~GHz, while the linewidth abruptly decreased and the integral generation power increased [Fig.~\ref{fig2}(b) and (c)]. In contrast to the  propagating spin wave mode at $I<I_t$, the oscillation exhibited a red shift with increasing $I>I_t$. An additional small peak at $f=7.4$~GHz close to $f_{FMR}$ appeared in the spectrum at $I>14$~mA, correlated with the decrease of the low-frequency peak intensity. Similar spectral features were previously observed in SCAO with in-plane magnetic anisotropy~\cite{Demidov_SHO,rhliu} and the conventional multilayer STNO~\cite{bonetti}. The low-frequency peak has been identified with the self-localized spin wave "bullet" oscillation mode~\cite{slavinprl}, its magnetization configuration resembles a droplet with almost uniform magnetization distribution in the domain wall (DW)~\cite{SM}. The peak close to $f_{FMR}$ is likely associated with a quasi-propagating mode that becomes weakly localized due to the Oersted field, which is opposite to $H$~\cite{hoeffer2}.

\begin{figure}[htbp]
\centering
\includegraphics[width=0.40\textwidth]{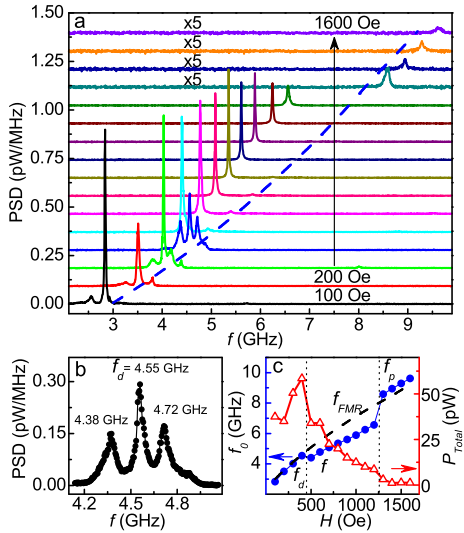}
\caption{(Color online). Dependence of the microwave generation characteristics on $H$ at $I=13$~mA. (a) Spectra obtained at $H$ between $100$~Oe and $1.6$~kOe increased in $100$~Oe steps. (b) Spectrum obtained at $H$=400~Oe exhibits sidebands around the main peak at $f_d$=4.55 GHz. (c) Dependence of the central generation frequency (solid symbols), $f_{FMR}$ (dashed curve), and total integral intensity (open symbols) on $H$.} \label{fig3}
\end{figure}

Since the static magnetization configuration of the Pt/[Co/Ni] multilayer depends on the in-plane field [see the discussion of Fig.~\ref{fig1}(a)], one can also expect it to exhibit different field-dependent dynamical regimes. Indeed, the oscillation peak abruptly develops sidebands below $H=500$~Oe, while at $H=1.2$~kOe the oscillation frequency jumps from below to above $f_{FMR}$ [Fig.~\ref{fig3}]. This latter transition is similar to the transition from the bullet mode to the quasi-linear propagating spin wave mode observed when decreasing $I$ at a constant $H$ [Fig.~\ref{fig2}]. It can be attributed to the increase of $I_t$ with $H$, indicating that the propagating mode is stabilized by the field. Therefore, our data show that one can control the dynamical mode in PMA-SCAO {\it either} by varying  field at a fixed current, {\it or} by varying current at a fixed field. We note that in contrast to the conventional multilayer STNO with in-plane anisotropy~\cite{dumas}, one mode disappears and another mode appears precisely at the transition driven either by current [Fig.~\ref{fig2}] or by field [Fig.~\ref{fig3}]. These behaviors suggest that the oscillation is always excited in the same spatial region, and that different oscillation modes of the same region are mutually exclusive.

The spectral transition observed when decreasing $H$ below $500$~Oe is characterized by an abrupt onset of the modulation sidebands separated from the main peak by multiples of $\Delta f\approx 170$~MHz [Fig.~\ref{fig3}(b)]. In addition, the central generation frequency shifts close to $f_{FMR}$, the linewidth of the main peak decreases, and the integral emitted microwave power dramatically increases [Fig.~\ref{fig3}(c)]. The observed spectral characteristics are reminiscent of the spin wave ``droplet" soliton observed in a conventional multilayer STNO with PMA of the active magnetic layer, at large fields normal to the film plane~\cite{mohseni}. The ``droplet" consists of a central region with magnetization direction opposite to the surrounding magnetic film separated from the latter by a region of precessing magnetization, producing a spatial magnetization distribution that resembles a nanoscale magnetic bubble. We note that the observed onset of the modulation peaks at $H<500$~Oe coincides with the opening of the hysteresis loop associated with the stabilization of the magnetic bubble domains in the [Co/Ni] film [see the discussion of Fig.~\ref{fig1}(a)]. These features were similar in all tested samples and reproducible at $H>100$Oe, indicating that the magnetization configuration of the active device area is dominated by the current-induced effects. At $H<100$~Oe, we observed a hysteresis of spectral characteristics consistent with history-dependent pinning of the magnetic bubbles on imperfections.

\begin{figure}[htbp]
\centering
\includegraphics[width=0.35\textwidth]{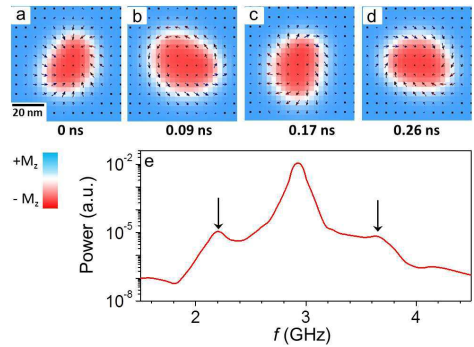}
\caption{(Color online). (a)-(d). Time sequences of out-of plane (in color) and in-plane (vector) magnetization components obtained from micromagnetic simulations at current $I=5$ mA and field $H=300$~Oe. (e) The power spectrum of the averaged in-plane projection of magnetization along the current direction. Two arrows indicate sideband peaks. The material parameters used in the simulations are the exchange stiffness $A=10$ pJ/m, the saturation magnetization $M_s=850$~kA/m, the perpendicular anisotropy $K_u=0.65$~MJ/m$^3$, the damping constant $0.02$, and the spin Hall angle $0.05$.}
\label{fig4}
\end{figure}

The similarity of the spectral features observed at small $H$ to those of the ``droplet" solitons, and the correlation of this dynamical regime with the bubble domain state indicate that these features are associated with the dynamics of a nanoscale magnetic bubble trapped in the active region of SCAO. The results of the micromagnetic simulations performed using the OOMMF code~\cite{oommf} support this interpretation and provide insight into the spatial, dynamical and topological characteristics of this state. The contributions from the demagnetizing field, exchange interaction, perpendicular anisotropy, the Zeeman field, the in-plane component of the Oersted field of the current~\cite{comsol}, and the effect of ST were taken into account. The value of magnetic anisotropy $K_u$ and the saturation magnetization $M_{sat}$ of the [Co/Ni] film were determined separately from the out-of-plane FMR data and vibrating sample magnetometry.

The simulations performed at a small in-plane field $H=300$~Oe show that a nanoscale magnetic bubble is stabilized in the active region of SCAO by the effects of current. The snapshots of the instantaneous magnetization distribution [Fig.~\ref{fig4}(a)-(d)] show that the bubble consists of a $40$~nm inverted core separated from the surrounding film by a $10$~nm-wide dynamical region that periodically evolves between configurations equivalent to a Bloch- or N\`{e}el-type DW. Since the magnetization in this region goes between `up' and `down' directions through the in-plane orientation, the ST due to the in-plane polarized spin current is large. The efficiency of ST-induced excitation is consistent with large microwave power observed in our measurements [Fig.~\ref{fig3}(c)]. The magnetization in the DW region precesses in the clockwise direction [top view], resulting in periodic changes of its helicity $\gamma$ from $\pi$ in panel (a) to $\pi/2$ in (b), $0$ in (c), and $-\pi/2$ in (d)~\cite{nagaosa}. The simulation also shows that the core region periodically shrinks for $\gamma=0,\pi$ [panels (a),(c)] and expands for $\gamma=\pm\pi/2$ [panels (b), (d)], and simultaneously exhibits a gyrotropic motion not synchronized with the precession. To simulate the microwave AMR signal produced by this dynamics, the Fourier spectrum of the spatially averaged projection of magnetization on the direction of the current flow was calculated. The spectrum consists of a large central peak associated with the DW precession, and sidebands caused by the gyration of the core [Fig.~\ref{fig4}(e)]. These spectral characteristics are in a qualitative agreement with the experimental observations [Fig.~\ref{fig3}].

To determine the topological properties of the dynamical state, its topological number $N$ was calculated by integrating the topological density of each magnetization snapshot over the plane, $N=\frac{1}{4\pi}\int\int (\partial_x\mathbf{m} \times\partial_y\mathbf{m})\cdot \mathbf{m} dxdy$, giving a time-independent value $N\simeq1$~\cite{Braun}. Therefore, the nanoscale bubble is topologically equivalent to a skyrmion trapped by the effects of current in the active device area, whose magnetic configuration dynamically evolves, under the influence of ST, among the instantaneous states characterized by different values of the skyrmion helicity $\gamma$~\cite{nagaosa}. We note that the skyrmion core gyration is consistent with the theoretical predictions~\cite{skew, moutafis,makhfudz}. The spectral characteristics and the spatial magnetization maps calculated for the dynamical skyrmion state resemble those of thee ``droplet" soliton~\cite{mohseni}. However, their topological properties are different: the latter is a non-topological dynamical state ($N=0$), and requires a larger field  and ST to sustain due to its dissipative properties~\cite{zhou,hoefer1,mohseni}. In contrast, the dynamical skyrmion is topologically nontrivial ($N=1$) and can be sustained by only spin current at zero field. Furthermore, a static skyrmion ($\gamma = \pi$ or 0) can become stabile at $I=0$ in utltrathin magnetic filmes with a sufficiently large DMI, which which can be achieved by magnetic interface engineering~\cite{romming}. This static state is expected to evolve from the current-induced dynamical  skyrmion when the current is removed~\cite{zhou}.

The dynamical skyrmion state obtained in our simulations became unstable at $H>300$~Oe or $I>5$~mA, and stripe-like domains or a quasi-uniform state were formed instead. In contrast, the experimentally observed spectral features identified with the skyrmion were observed up to $H=400$~Oe and $I=15$~mA. This discrepancy is likely caused by the limitations of our simulation. First, atomically sharp electrodes were assumed in the simulations, while the actual electrodes were rounded due to the finite resolution of e-beam lithography and the limitations of the sputtering process. As a consequence, the simulated current distribution is more localized, resulting in lower currents needed to nucleate and stabilize a dynamical skyrmion state. It is also possible that the DMI at the Pt/[Co/Ni] interface neglected in our simulation stabilizes the DW spin structure of the skyrmion~\cite{bode,chen}. In addition, the current-induced field-like torque neglected in the simulation may provide a non-neglibible contribution stabilizing the skyrmion core~\cite{liuprb}. Despite some quantitative discrepancies, our simulations confirm the possibility to create a stable dynamical skyrmion state in the SCAO~\cite{SM}.

To summarize, we have observed several distinct dynamical regimes of the SCAO based on a magnetic multilayer with PMA. By varying the field and current, both localized and propagating-mode dynamical regimes were obtained. At small fields, the spectral characteristics resembled the ``droplet" soliton. Experimental analysis and simulations showed that a nanoscale magnetic bubble topologically equivalent to the skyrmion becomes trapped in the active device region. In contrast to the non-topological droplet, the observed dynamical skyrmion is a topologically nontrivial structure and in films with a suficiently large DMI can remain stable in the absence of current and field, making this state a viable candidate for topologically stable spin-torque device applications~\cite{romming,nagaosa,locatelli}. Our results and analysis indicate that the spin current also induces a well-defined precession of the skyrmion's boundary magnetization, resulting in a periodic variation of the skyrmion helicity from $-\pi$ to $\pi$. Thus, it may be possible to use spin current not only to generate skyrmions, but also to dynamically control their helicity.

This work was supported by the NSF Grants No. ECCS-1305586, and No. DMR-1218414.


\begin{references}

\bibitem{slon1}
J. C. Slonczewski, \Journal{J. Magn. Magn. Mater.} {159}{L1} {1996}; \Journal{J. Magn. Magn. Mater.}{195}{L261} {1999}.

\bibitem{berger}
L. Berger, \Journal{Phys. Rev. B}{54}{9353}{1996}; \Journal{J. Appl. Phys.}{90}{4632} {2001}.

\bibitem{tsoiprl} M. Tsoi, A.G.M. Jansen, J. Bass, W.C. Chiang, M. Seck, V. Tsoi and P. Wyder, \Journal{Phys. Rev. Lett.}{80}{4281}{1998};\Journal{Phys. Rev. Lett.}{81}{493} {1998}.

\bibitem{cornellorig} J.A. Katine, F.J. Albert, R.A. Buhrman, E.B. Myers, and D.C. Ralph, \Journal{Phys. Rev. Lett.}{84}{3149}{2000}.

\bibitem{Ando_SHE}
K. Ando, S. Takahashi, K. Harii, K. Sasage, J. Ieda, S. Maekawa, and E. Saitoh, \Journal{Phys. Rev. Lett.}{101}{036601}{2008}.

\bibitem{wang}
Z.H. Wang, Y.Y. Sun, M.Z. Wu, V. Tiberkevich, and A. Slavin, \Journal{Phys. Rev. Lett.} {107}{146602} {2011}.

\bibitem{liuprl}
L. Liu, O.J. Lee, T.J. Gudmundsen, D.C. Ralph,and R.A. Buhrman, \Journal{Phys. Rev. Lett.} {109} {096602} {2012}.

\bibitem{miron}
I.M. Miron, K. Garello, G. Gaudin, P.J. Zermatten, M.V. Costache, S. Auffret, S. Bandiera, B. Rodmacq, A. Schuhl, and P. Gambardella, \Journal{Nature (London)} {476}{189} {2011}.


\bibitem{rashba}
Y. A. Bychkov and E. I. Rashba, \Journal{J. Phys. C} {17} {6039} {1984};
G. Dresselhaus, \Journal{Phys. Rev.} {100}{580} {1955}.

\bibitem{dmi}
T. Moriya,\Journal{Phys. Rev.} {120} {91} {1960}; I. E. Dzyaloshinskii, \Journal{Sov. Phys.} {5} {1259} {1957}.

\bibitem{Braun}
H.-B. Braun, \Journal{Adv. Phys.} {61} {1} {2012}.

\bibitem{obler}
U.K. R\"{o}{\ss}ler, A. N. Bogdanov and C. Pfleiderer, \Journal{Nature (London)} {442}{797} {2006}.


\bibitem{sampaio}
J. Sampaio, V. Cros, S. Rohart, A. Thiaville and A. Fert, \Journal{Nat. Nanotechnol.} {8} {839} {2013}.

\bibitem{zhou}
Y. Zhou, E. Iacocca, R. K. Dumas, F. C. Zhang, and J. {\AA}kerman, arXiv:1404.3281.

\bibitem{bode}
M. Bode, M. Heide, K. von Bergmann, P. Ferriani, S. Heinze, G. Bihlmayer, A. Kubetzka, O. Pietzsch, S. Bl\"{u}gel and R. Wiesendanger, \Journal{Nature (London)} {447} {190} {2007}.

\bibitem{romming}
N. Romming, C. Hanneken, M. Menzel, J. E. Bickel, B. Wolter, K. von Bergmann, A. Kubetzka, R. Wiesendanger, \Journal{Science} {341} {636} {2013}.


\bibitem{hoefer1}
M.A. Hoefer, T.J. Silva, and M.W. Keller, \Journal{Phys. Rev. B} {82}{054432} {2010}.

\bibitem{mohseni}
S.M. Mohseni, S.R. Sani, J. Persson, T.N.A. Nguyen, S. Chung, Y. Pogoryelov, P.K. Muduli, E. Iacocca, A. Eklund, R.K. Dumas, S. Bonetti, A. Deac, M. A.
Hoefer, and J.{\AA}kerman, \Journal{Science} {339}{1295} {2013}.

\bibitem{bubble}The nanoscale magnetic bubble with a well-defined chiral structure of the domain wall is characterized by the unity topological number and is topologically equivalent to the skyrmion.

\bibitem{Daalderop} G.H.O.Daalderop, P.J. Kelly, and F.J.A. den Broeder, \Journal{Phys. Rev. Lett.}{68}{682}{1992}.

\bibitem{simububble} Calculations performed with OOMMF micromagnetic simulation package using the experimentally determined magnetic properties of the film show that bubble domains are stable below $H_{in}\sim$ 300 Oe.

\bibitem{gubbiotti}
G. Gubbiotti, G. Carlotti, S. Tacchi, M. Madami, T. Ono, T. Koyama, D. Chiba, F. Casoli, and M.G. Pini, \Journal{Phys. Rev. B} {86} {014401} {2012}.

\bibitem{chen}
G. Chen, T. P. Ma,	A. T. N'Diaye,	H. Y. Kwon,	C. Won,	Y. Z. Wu and A. K. Schmid, \Journal{Nat. Comm.} {4}{2671} {2013}.

\bibitem{mcguire}
T.R. Mcguier and R.I. Potter, \Journal{IEEE Trans. Mag.} {4}  {1018-1038} {1975}.

\bibitem{mosendz}
O. Mosendz, J. E. Pearson, F.Y. Fradin, G.E.W. Bauer, S.D. Bader, and A. Hoffmann, \Journal{Phys. Rev. Lett.} {104}{046601} {2010}.

\bibitem{Iac} We have separately checked that varying the excitation current does not significantly affect the ST-FMR results.

\bibitem{Demidov_SHO} V.E. Demidov, S. Urazhdin, H. Ulrichs, V. Tiberkevich, A. Slavin, D. Baither, G. Schmitz, and S. O. Demokritov, \Journal{Nat. Mater.} {11}{1028}{2012}.

\bibitem{rhliu}
R.H. Liu, W.L. Lim, and S. Urazhdin, \Journal{Phys. Rev. Lett.} {110}{147601} {2013}

\bibitem{bonetti}
S. Bonetti, V. Tiberkevich, G. Consolo, G. Finocchio, P. Muduli, F. Mancoff, A. Slavin, and J. {\AA}kerman, \Journal{Phys. Rev. Lett.} {105}{217204}{2010}.

\bibitem{slavinprl}
A. Slavin and V. Tiberkevich, \Journal{Phys. Rev. Lett.} {95}{237201}{2005}.

\bibitem{SM}
See Supplemental Material at [url] for additional simulations and discussion of spin current nano-oscillator with perpendicular
magnetic anisotropy, which includes Refs[32-35].

\bibitem{S2}
S. Rohart and A. Thiaville, \Journal{Phys. Rev. B} {88} {184422} {2013}.

\bibitem{S4}
T. H. O'Dell, \Journal{Rep. Prog. Phys.} {49} {589} {1986}.

\bibitem{S6}
S. Emori, U. Bauer, S.M. Ahn, E. Martinez, and G. S. D. Beach, \Journal{Nat. Mater.} {12} {611} {2013}.

\bibitem{S7}
P. P. J. Haazen, E. Mur\`{e}, J. H. Franken, R. Lavrijsen, H. J. M. Swagten, and B. Koopmans, \Journal{Nat. Mater.} {12} {299} {2013}.


\bibitem{hoeffer2}
M. A. Hoefer, T. J. Silva, and M. D. Stiles, \Journal{Phys. Rev. B} {77} {144401} {2008}.


\bibitem{dumas}
R.K. Dumas, E. Iacocca, S. Bonetti, S.R. Sani, S. M. Mohseni, A. Eklund, J. Persson, O. Heinonen, and J. {\AA}kerman, \Journal{Phys. Rev. Lett.} {110} {257202}{2013}.


\bibitem{oommf}
M.J. Donahue and D.G. Porter, OOMMF User's Guide, Interagency Report NISTIR 6376, NIST Gaithersburg, MD (1999); \href{http://math.nist.gov/oommf}{http://math.nist.gov/oommf}.

\bibitem{comsol} The current density distribution and the in-plane component of the Oersted field were calculated by the commercially available {\it comsol} software.

\bibitem{nagaosa} N. Nagaosa and Y. Tokura, \Journal{Nat. Nanotech.}{8}{899}{2013}.

\bibitem{skew}
N. Papanicolaou and T. N. Tomaras, \Journal{Nucl. Phys. B} {360} {425}{1991};
S. Komineas and N. Papanicolaou, \Journal{Physica D} {99} {81} {1996}.

\bibitem{moutafis}
C. Moutafis, S. Komineas, and J. A. C. Bland, \Journal{Phys. Rev. B} {79} {224429} {2009}.

\bibitem{makhfudz}
I. Makhfudz, B. Kr\"{u}ger, and O. Tchernyshyov, \Journal{Phys. Rev. Lett.} {109} {217201} {2012}.

\bibitem{liuprb}
R. H. Liu, W. L. Lim, and S. Urazhdin, \Journal {Phys. Rev. B} {89} {220409(R)}{2014}.

\bibitem{locatelli}
 N. Locatelli, V. Cros, and J. Grollier, \Journal{Nat. Mater.} {13}{11}{2014}.


\noindent
\end{references}
\end{document}